\documentclass[aip,reprint]{revtex4-1}
\usepackage{graphicx}
\graphicspath{ {./pix/} }
\usepackage{epstopdf}
\usepackage{color}
\usepackage{bm}
\usepackage{color,soul}
\usepackage[colorlinks=true,linkcolor=blue]{hyperref}
\expandafter\ifx\csname package@font\endcsname\relax\else
 \expandafter\expandafter
 \expandafter\usepackage
 \expandafter\expandafter
 \expandafter{\csname package@font\endcsname}%
\fi
\begin{document}

\title{Broadband Filters for Abatement of Spontaneous Emission in Circuit Quantum Electrodynamics}

\author{Nicholas T. Bronn}
\email{ntbronn@us.ibm.com}
\affiliation{\mbox{IBM TJ Watson Research Center, 1101 Kitchawan Road, Yorktown Heights, NY 10598, USA}}

\author{Yanbing Liu}
\affiliation{\mbox{Department of Electrical Engineering, Princeton University, Princeton, NJ 08544,~USA}}

\author{Jared B. Hertzberg}
\author{Antonio D. C\'orcoles}
\affiliation{\mbox{IBM TJ Watson Research Center, 1101 Kitchawan Road, Yorktown Heights, NY 10598, USA}}

\author{Andrew A. Houck}
\affiliation{\mbox{Department of Electrical Engineering, Princeton University, Princeton, NJ 08544,~USA}}

\author{Jay M. Gambetta}
\author{Jerry M. Chow}
\affiliation{\mbox{IBM TJ Watson Research Center, 1101 Kitchawan Road, Yorktown Heights, NY 10598, USA}}

\date{\today}


\begin{abstract}

The ability to perform fast, high-fidelity readout of quantum bits (qubits) is essential to the goal of building a quantum computer. However, coupling a fast measurement channel to a superconducting qubit typically also speeds up its relaxation via spontaneous emission. Here we use impedance engineering to design a filter by which photons may easily leave the resonator at the cavity frequency but not at the qubit frequency. We implement this broadband filter in both an on-chip and off-chip configuration.

\end{abstract}

\maketitle


Superconducting qubits have become strong candidates for implementing fault-tolerant quantum computing~\cite{Riste2015,Kelly2015,Corcoles2015} and digital quantum simulations~\cite{Heras2014,Barends2015}. Recent progress has been driven by improved coherence times~\cite{Paik2011,Chang2013a}, gate fidelities~\cite{Chow2011,Barends2014}, and readout fidelities~\cite{Siddiqi2004,Riste2012,Liu2014}. The demonstration of a fault-tolerant logical qubit will require both further improvements in these areas and also the engineering of architectures for inter-connected networks of superconducting qubits~\cite{Chow2014}.

One critical area of exploration for larger networks of superconducting devices is the complementary pursuit of fast, high-fidelity readout and suppression of qubit decay. Currently, low qubit decay rates are made possible by coupling the superconducting qubit to a microwave resonator in the circuit quantum electrodynamics (cQED) architecture~\cite{Blais2004,Wallraff2004}. The microwave resonators protect the qubits from spontaneous emission into the environment. With the appropriate coupling, they also permit a quantum non-demolition readout of the qubit state. However, suppression of the qubit decay rate comes at a cost of the readout rate. To speed up the readout, a number of methods have been proposed for dispersive filtering, where radiation at the qubit frequency is filtered and that at the resonator frequency is passed~\cite{Reed2010, Jeffrey2014, Kelly2015, Bronn2015}. Among these proposals, large bandwidth filters with the possibility of off-device integration have been absent. In this Letter we demonstrate the suppression of qubit decay rates with a broadband stepped impedance Purcell filter (SIPF) in both on- and off-chip configurations, while maintaining the ability to perform fast, high-fidelity readout.


A qubit coupled to the environment suffers relaxation based on the admittance (reciprocal impedance) $Y(\omega_q)$ at its transition frequency $\omega_q$. This dependence of spontaneous emission on the coupled electromagnetic environment is known as the Purcell effect~\cite{Purcell1946}, and is a key factor used to either enhance or abate qubit relaxation. Approximating a transmon qubit as a harmonic oscillator, its lifetime is
\begin{eqnarray} \label{eq:t1}
T_1 = C_\Sigma/{\rm Re}[Y(\omega_q)],
\end{eqnarray}
where $C_\Sigma$ is the sum of the shunting and Josephson capacitances~\cite{Koch2007}. Initial experiments coupling superconducting qubits to the external environment (i.e. measurement and control instruments) through coupling capacitors and tunnel junctions~\cite{Nakamura1999}  did not provide sufficient protection and thus yielded large decay rates $\gamma_1 = T_1^{-1}$. Placing the qubit inside a resonator modifies the available decay channels and hence $Y(\omega_q)$, which forms the basis of cQED~\cite{Blais2004,Wallraff2004}. In particular, admittance to the external environment decreases as the qubit-resonator frequency detuning $\Delta$ increases. Coupling to higher order harmonics of the resonator may be minimized if the qubit frequency is lower than the fundamental mode resonance~\cite{Houck2008}. 

Within the cQED architecture, additional qubit protection may be achieved by incorporating a Purcell filter. To suppress qubit decay (the Purcell effect) the filter presents an impedance mismatch at the qubit frequency, while permitting measurement by matching the 50~$\Omega$  environment at the readout resonator frequency. Early demonstrations of such filters employed a transmission line stub one-quarter wavelength long at the qubit frequency, terminated in an open circuit and placed on the output port of a qubit-resonator device~\cite{Reed2010}. This filter prevents qubit decay by appearing as a short circuit at the qubit frequency. A similar single-pole filter can be realized by tuning a stray capacitance between the qubit and external environment~\cite{Bronn2015}. Other Purcell filters instead use a low-$Q$ resonator to multiplex the readout resonators of multiple qubits \cite{Jeffrey2014,Kelly2015,Sete2015}. Since the qubits are detuned from the resonator frequencies in cQED, the low-$Q$ resonator further suppresses qubit decay. More complicated structures such as a multimode bandpass filters have also been used to suppress off-resonant coupling between qubits~\cite{McKay2015}.

Fixed-frequency qubits used in our typical experiments~\cite{Chow2014,Corcoles2015} are not affected by $1/f$ flux noise, however their frequencies are very sensitive to fabrication conditions~\cite{Chow2015} and it could be difficult to align them with a narrow band such as the $\lambda/4$ stub filter. Although low-$Q$ resonator bandpass filters do not present this problem of frequency alignment and are also conducive to multiplexed readout, they are much less attenuating than narrowband notch filters~\cite{Jeffrey2014, Kelly2015}. These resonant designs also filter low-frequency signals, prohibiting their use with DC and fast flux bias lines.


\begin{figure}[htbp]
	\centering
	\includegraphics[width=3.375in]{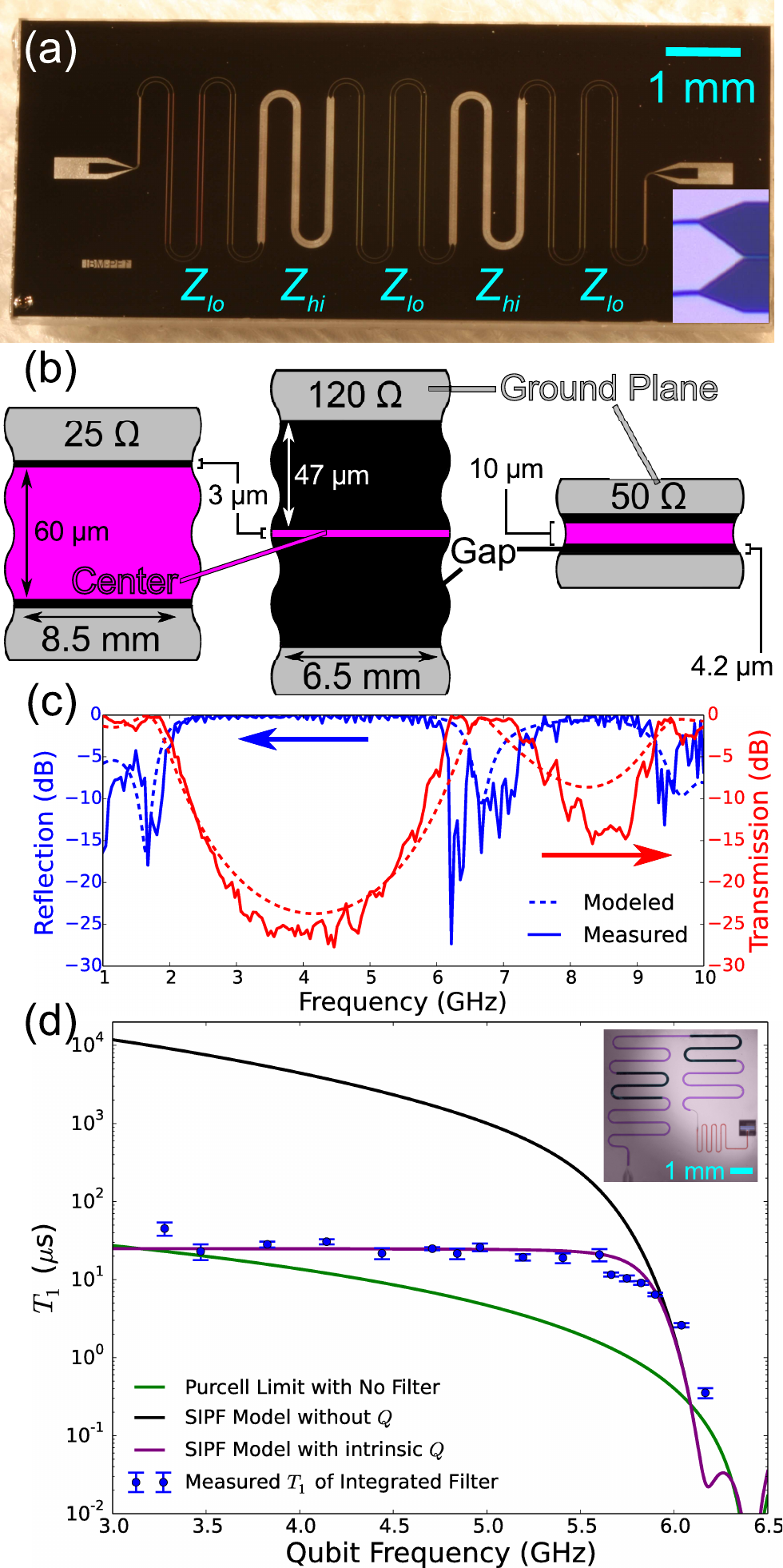}
	\caption{\label{fig1} (a) SIPF as fabricated on sapphire. Inset: the transition between low-impedance (left) and high-impedance (right) CPW sections takes place over $\sim$~40~$\mu$m. (b) Schematic of $Z_{\rm lo}$, $Z_{\rm hi}$, and $Z_{\rm env} = 50$~$\Omega$ (as a reference) CPW sections detailing the length, center (purple) and gap (black) widths and the ground plane in gray. (c) Filter reflection and transmission, both modeled and measured. The ripples are likely due to impedance mismatches in the sample package. (d) Qubit lifetime, both measured and predicted from an estimate of the Purcell effect with and without the filter and intrinsic qubit loss. The dip in the predicted $T_1$ around 6.2~GHz arises from the inclusion of a wirebond as a 2~nH inductor. Inset: false-colored micrograph of the integrated SIPF device on sapphire showing SIPF (purple/black), resonator (red), and qubit in the dark rectangular pocket.}
\end{figure}

Here we present the stepped impedance Purcell filter (SIPF), consisting of alternating sections of high and low impedance coplanar waveguide (CPW) transmission lines \cite{Pozar2012}, which features a wide stopband easily encompassing the natural spread in fixed-frequency qubits. While typically used as a lowpass filter, here the SIPF design is implemented as a dual bandstop/bandpass filter. Fig.~\ref{fig1}a-b show the filter device and geometry of the five-section SIPF: alternating $Z_{\rm lo}$~$=$~$25$~$\Omega$ and $Z_{\rm hi}$~$=$~$120$~$\Omega$ CPW sections with the $Z_{\rm lo}$ sections forming the ends. The modeled and measured transmission and reflection characteristics of our filter are shown in Fig.~\ref{fig1}c, featuring a wide stopband from $2$~$-$~$6$~GHz and passband around 6.5~GHz. Additionally, the filter is `flat' enough to pass qubit control pulses with the maximum attenuation around 25 dB. The filter also possesses a DC passband, which would allow for fast flux biasing, so that a single filter and single line could be used for both measurement and control lines.
 
 The SIPF is integrated into the feedline of a tunable transmon qubit device for measurement in reflection at the resonator frequency of 6.42~GHz, matching the passband of the SIPF. The device is fabricated from 200~nm of sputtered niobium dry etched using SF$_6$ on a 7~$\times$~7~mm$^2$ sapphire substrate with a thickness of 530~$\mu$m, as seen in the inset of Fig.~\ref{fig1}d. Qubit lifetimes were measured at the base temperature of a dilution refrigerator ($\sim~10$~mK), and plotted in Fig.~\ref{fig1}d, along with predictions for the Purcell limit with no filter and for the SIPF with and without intrinsic $Q$ of the qubit. The models are based on Eq.~\ref{eq:t1}, and good agreement is found for the SIPF model with intrinsic $Q \approx 1\,000\,000$. A clear enhancement of qubit lifetime is thereby demonstrated, and in particular, around 5 GHz, where our experiments with fixed-frequency qubits are performed~\cite{Chow2014,Corcoles2015}, we measure $T_1$~$=$~$26$~$\mu$s, whereas the predicted lifetime without the filter is 5~$\mu$s, based on the measured linewidth $\kappa/2\pi$~$\approx$~$7$~MHz. As a figure of merit (FOM), we calculate a lifetime bandwidth product by integrating the $T_1$ bound of the filter models over the frequency spread of fixed-frequency qubits. Though designed for frequency of 5~GHz, fabrication conditions give a frequency spread of approximately 500~MHz~\cite{Chow2015}. We find an FOM for the SIPF of $16\,000$, compared with $3\,700$ for the low-$Q$ bandpass filter~\cite{Jeffrey2014}, while the FOM diverges for the quarter-wave stub~\cite{Reed2010}.

An intuitive understanding of the SIPF operation may be gleaned by considering the propagation through a SIPF having an infinite number of sections. Using $ABCD$ matrices \cite{Pozar2012} to calculate the dispersion relation, the filter cutoffs are determined by
\begin{eqnarray*} \label{eq:dispersion}
\bigg|2\cos k_c\ell_{\rm lo} \cos k_c\ell_{\rm hi} - (\alpha+1/\alpha) \sin k_c\ell_{\rm lo} \sin k_c\ell_{\rm hi}\bigg| = 2
\end{eqnarray*}
where $k_c = 2\pi/\lambda_c = \omega_c/v_p$ is the cutoff wavenumber given phase velocity $v_p$ and $\alpha = Z_{\rm hi}/Z_{\rm lo}$ is the impedance asymmetry of the filter. Here it is seen that the lengths $\ell$ of the sections define the $\lambda/2$ resonant frequencies and hence the passbands of the filter, while the rejection bandwidth increases with $\alpha$. In this approximate model, the parameters of our filter yield a stopband from 2.6 to 5.7~GHz and a passband centered at 6.6~GHz, in good agreement with Fig.~\ref{fig1}c. Applying the technique of $ABCD$ matrices numerically to a finite number of sections, one can easily observe increasing bandstop insertion loss by increasing the impedance asymmetry and/or the number of sections.

\begin{figure}[!tbp]
	\centering
	\includegraphics[width=3.375in]{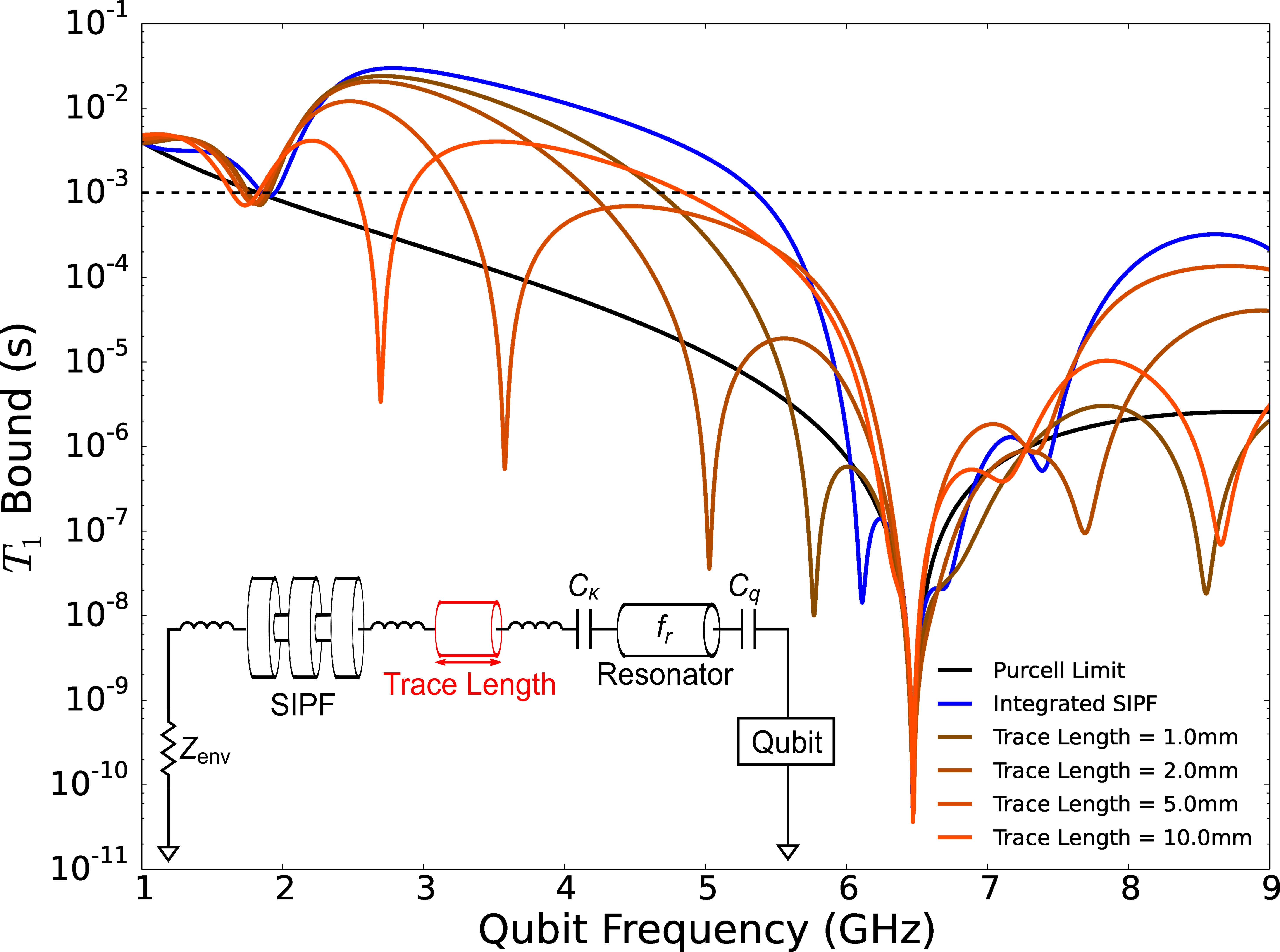}
	\caption{\label{fig2} Predicted $T_1$ bound as a function of qubit frequency for the ``standalone'' configuration of the SIPF, showing a ``dip'' below the resonator frequency $f_r$~$\sim$~$6.5$~GHz that depends on the length of the trace, as compared with the $T_1$ bound for the integrated SIPF. The dashed line indicates a lifetime of 1 ms. Inset: The copper trace is modeled as a 50 $\Omega$ transmission line connected to the qubit device and SIPF by wirebonds acting as 2~nH inductors. The capacitors $C_\kappa$ and $C_q$ define the resonator decay rate $\kappa$ and qubit-resonator coupling, respectively.}
\end{figure}

As die size grows, spurious package modes may lead to additional relaxation channels that couple adversely to the qubits \cite{Srinivasan2015}. With a total length of 37.5 mm, SIPFs are far too large to be integrated on-chip with multiple qubits while avoiding these spurious modes. Instead, it is desirable to locate them off-chip in order to keep the qubit chip small. A further benefit of having the SIPFs off-chip is that qubit devices may be replaced without replacing the SIPF itself, a modular approach that may enable, for example,  multiboard integration using interposer technology~ \cite{Colless2012}. We envision a multi-qubit architecture in which ``standalone'' filters fabricated on separate chips are mounted on the same printed circuit board (PCB) package. Our device packaging~\cite{Chow2014,Corcoles2015} must include a 50~$\Omega$ PCB signal trace between the readout resonator and SIPF, connected to both the readout resonator and the SIPF input port via wirebonds.

The effect of this signal trace on $T_1$ bound was again modeled using Eq.~\ref{eq:t1} with the circuit depicted in the inset of Fig.~\ref{fig2}, and compared to that expected from the ``integrated'' SIPF. Here a dielectric constant of 3.66 and loss tangent of 0.0127 were obtained from the datasheet for FR408 dielectric~\cite{FR408} and a resistance per unit length of 8.7~n$\Omega/\mu$m derived from Ref.~\citenum{Copper} for the copper signal trace. Additionally the aluminum wirebonds were modeled as 2 mm long loops acting as 2 nH inductors. As seen in Fig.~\ref{fig2}, the copper signal trace introduces an undesired dip in the $T_1$ bound at a frequency below the resonator frequency. Increasing the length of the signal trace reduces the `dip' frequency, but making it too long adds extra modes. A length of approximately 10 mm produces single dip in $T_1$ bound at about 2.7~GHz, well below the transition frequency of our fixed-frequency qubits.


\begin{figure}[tb]
	\centering
	\includegraphics[width=3.375in]{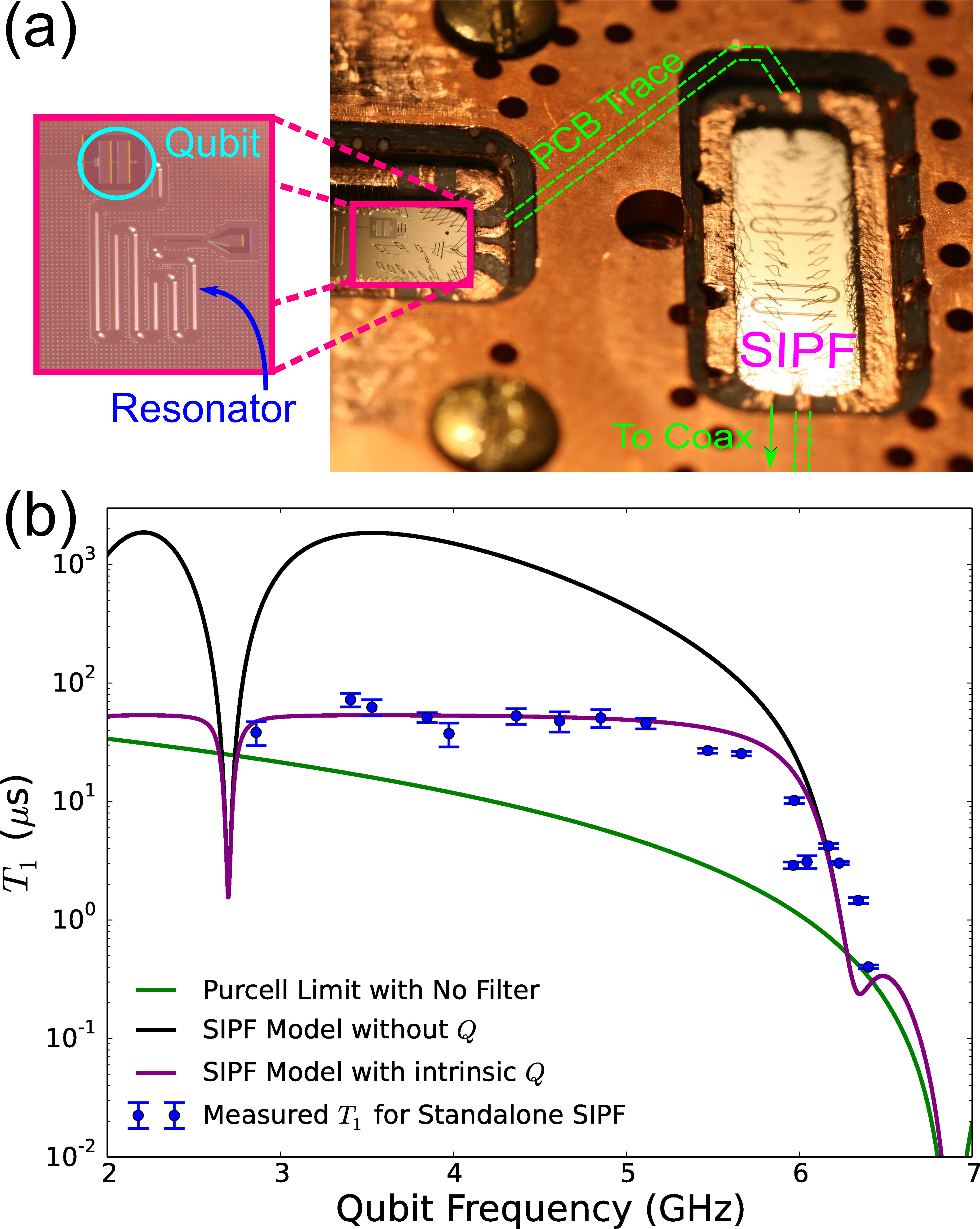}
	\caption{\label{fig3} (a) Sample package used to mount qubit/resonator device with standalone SIPF. The trace (in green) is buried in the signal plane of the PCB. (b) Measured and predicted qubit lifetime. The Purcell effect is estimated with and without the filter and intrinsic loss ($Q$~$\approx$~$2\,400\,000$) in the standalone configuration with 10 mm of copper trace between the qubit on silicon and SIPF on sapphire. The small dip just below the resonator frequency in the SIPF models is due to the appearance of the second mode.}
\end{figure}

The test measurement package consists of the 4~$\times$~8~mm$^2$ tunable qubit/resonator silicon chip and 4~$\times$~10~mm$^2$ sapphire filter chip back-mounted to a PCB by a copper push block. The PCB has windows milled out to accommodate the active area of the samples, as shown in Fig.~\ref{fig3}a. The PCB contains four copper layers separated by FR408 dielectric, three of which are ground, and a buried copper stripline as the signal trace. The PCB is further milled away to expose the ends of the signal traces for wirebonding to the qubit and filter. The buried signal trace is highlighted in green and has a length of approximately 10 mm. The chips' ground planes are in intimate thermal and electrical contact with the PCB ground. Measured qubit lifetime versus frequency is plotted in Fig.~\ref{fig3}b, where enhancement over the Purcell limit is again apparent. In particular, around 5~GHz, $T_1$~$=$~$45$~$\mu$s, whereas the Purcell limit without the filter is 5~$\mu$s (as $\kappa/2\pi$~$\approx$~$7$~MHz again). Predicted $T_1$'s for the SIPF are also shown with and without intrinsic losses of qubits on silicon ($Q~\approx~2\,400\,000$). Without this intrinsic loss a further improvement in $T_1$ up to two orders of magnitude is possible. Importantly, the existence of the PCB trace between qubit/resonator and filter does not limit performance, implying that SIPFs with PCB traces may be connected to each readout resonator in a multi-qubit device.


The ability to move the filter off-chip opens up the possibility of integrating the SIPF with other superconducting coplanar circuits. In particular, the integration of filtering with on-chip circulators~\cite{Kamal2011,Kerckhoff2015} and quantum-limited amplifiers~\cite{Castellanos-Beltran2008,Bergeal2010,Abdo2011,Vijay2011,Hatridge2013} could reduce the number of bulky components needed at the mixing chamber stage of the dilution refrigerator. For devices with a larger number of qubits and readout frequencies, SIPFs are naturally good complements to broadband, near quantum-limited amplifiers such as the traveling wave parametric amplifier \cite{Macklin2015}. Additionally, the close proximity between qubit and amplifier mitigates against loss in coax lines. Integrated packaging of this sort might lend itself well for future extensibility.


In conclusion, we have demonstrated a broadband filter that suppresses the superconducting qubit decay which arises from the strong coupling to the readout cavity. This filter is demonstrated in both an integrated on-chip configuration and off-chip in a standalone configuration in the PCB package. The wide stopband of this filter provides protection for a large range of qubit frequencies while allowing qubit control through the same line. Furthermore, this will allow for fast, high-fidelity readout when paired with quantum-limited amplifiers.

We thank M.B. Rothwell and G.A. Keefe for fabrication of qubit devices, Y.-K.-K. Fung, J.R. Rozen, and J. Rohrs for experimental contributions, and D.C. McKay, A. Mezzacapo, and F. Solgun for insightful discussions. We acknowledge support from IARPA under contract W911NF-10-1-0324. 

\bibliographystyle{aipnum4-1}
\bibliography{paper-puffy-t1}

\end{document}